\documentclass[a4paper,10pt,one side]{article}
\usepackage{graphicx} 
\usepackage{graphicx}
\usepackage{titlesec}
\usepackage[utf8]{inputenc}
\usepackage[hidelinks]{hyperref}
\usepackage{multirow}
\usepackage{textcomp, gensymb}
\usepackage{siunitx}
\usepackage{float}
\usepackage[table,xcdraw]{xcolor}
\usepackage{blindtext}
\usepackage{geometry}
 \usepackage{multicol}
\usepackage[english]{babel}
\usepackage{geometry}
\usepackage{siunitx}
\usepackage{amsmath}
\usepackage{amssymb}
\usepackage{multirow}
\usepackage{verbatim}
\usepackage{contour}
\contourlength{2pt}
\usepackage[european,straightvoltages,EFvoltages]{circuitikz}
\usepackage{mathtools}
\usepackage[labelfont=bf]{caption}
\usepackage{pgfplots}
\pgfplotsset{compat=1.17}
\usepgfplotslibrary{fillbetween}
\usepackage{float}
\usepackage{csvsimple}

\usepackage{etoolbox}
\makeatletter
\patchcmd{\@maketitle}{\vskip 2em}{\vspace*{0cm}}{}{}
\makeatother
\date{}
\usepackage{natbib}
\setcitestyle{authoryear,open={(},close={)}} 
\usepackage{url}  
\usepackage{hyperref}  
\hypersetup{
    colorlinks=true,      
    linkcolor=blue,      
    citecolor=blue,      
    urlcolor=blue        
}

\titleclass{\subsubsubsection}{straight}[\subsection]

\newcounter{subsubsubsection}[subsubsection]
\renewcommand\thesubsubsubsection{\thesubsubsection.\arabic{subsubsubsection}}

\titleformat{\subsubsubsection}
  {\normalfont\normalsize\bfseries}{\thesubsubsubsection}{1em}{}
\titlespacing*{\subsubsubsection}
{0pt}{3.25ex plus 1ex minus .2ex}{1.5ex plus .2ex}

\title{``CREATOR Case'': PMSM and IM Electric Machine Data for Validation and Benchmarking of Simulation and Modeling Approaches}

\author{Annette Mütze,
Kourosh Heidarikani,
Pawan Kumar Dhakal,
Roland Seebacher\\
Electric Drives and Power Electronic Systems Institute, Graz University of Technology\\[1em]
Sebastian Schöps\\
Computational Electromagnetics Group, Technical University of Darmstadt%
}

\begin{document}

\maketitle
\vspace{-1cm}
\section*{Abstract}
\textbf{Purpose}\;--\;This paper describes the complete sets of data of two different machines, a PMSM and an IM, that are made available to the public for modeling and simulation validation and benchmarking.\smallskip \newline 
\textbf{Design/methodology/approach}\;--\;For both machines, not only the complete sets of design parameters, i.e., motor geometry, electrical parameters, material properties, and winding schemes as well as the measured low-frequency equivalent circuit parameters are provided, but also comprehensive measurement results on six different drive cycles to allow for transient investigations.\smallskip \newline
\textbf{Findings}\;--\;The data packages provide all the required information in terms of design parameters and measurement results that facilitate modeling and simulation validation and benchmarking results for verification of different modeling approaches.\smallskip \newline
\textbf{Research limitations/implications}\;--\;The paper serves as the key reference user manual for the extensive and comprehensive sets of data made available. It is therefore recommended to follow up on the reading of the paper by a study of the data packages themselves.\smallskip \newline
\textbf{Originality/value}\;--\;To the authors' best knowledge, this is the first time that the complete sets of machine data of two different machines are published, allowing for benchmarking of modeling and simulation projects, and reproducibility and reusability following the FAIR data principle of analyses based on these data.\newline
\\ \textbf{Keywords}\;--\;Electric machines, design data, measurement results, modeling, reusability, simulation, verification, validation.\newline

\section{Motivation}
The tedious design process of electric machines, that involves different disciplines, e.g., electrical, mechanical, magnetic, and thermal aspects, and also increasingly dynamic analyses, calls for example cases for benchmarking of modeling and simulation approaches. Today, the literature only presents selected machine design-related data and shows the laboratory results of chosen parameters or performance criteria. As verification and validation of models has become increasingly important, so has the need for the availability of data to allow for verification and validation of models, as well as reproducibility and reusability of results \citep{oberkampf_verification_2010}. To this aim, and to the authors' best knowledge for the first time, we publish the complete sets of machine data for two different machines for such multifaceted analysis. These packages of data provide all the required information in terms of design parameters and measurement results to facilitate their use as validation and benchmarking tool for different modeling approaches, beyond  the specific quantitative results as obtained from singular analyses. For example, \citet{bergfried_thermal_2023} have investigated the specific question of thermal modeling and simulation based on our data.

\section{Introduction}
Electric machines, as key players in energy conversion, have been known for more than a century. The recent technological advancements in electric drives and power electronic systems, together with new materials and manufacturing techniques and increasingly transient performance requirements, provide plenty of opportunities for future innovations, but also  increase the demands on the modeling and simulation tools, e.g., \citet{ahn_multiobjective_2023, hwang_coupled_2021}.
Current electric machine design procedures usually start with an expert's choice for a particular machine type and topology or already-developed design choices, e.g., \citet{barcaro_design_2013,6739386,mese_design_2016}. Optimization is then typically based on selected parameters and steady-state operating points, e.g., \citet{rimpas_comparative_2023,gobbi_traction_2024, li_analysis_2017}.  However, exploiting the full potential of electric machines in the future requires more powerful approaches to account for the different design possibilities, parameters, and criteria. We provide all data relevant for the comprehensive modeling of two machines, as well as measurement results for both selected steady-state operating points and three different drive cycles, to allow for code validation or benchmarking, e.g., as a baseline for innovative shape or topology optimization techniques. As a matter of fact, well-prepared experimental data is incredibly valuable for the validation of simulation models and available data can underpin the reliability and reproducibility of the results for the community \citep{weinper_2021}. 
We thereby also contribute to increasing availability of results in the field of electric machines that follow the FAIR (Findability, Accessibility, Interoperability, and Reusability) principles, which enables transparency and reproducibility of research knowledge, an acronym introduced almost a decade ago by a consortium of scientists and organizations \citep{wilkinson_2016}.
In contrast to the reporting of selected results without context, e.g., on data science competition platforms \citet{dataser_synchronous_nodate,dataset_experiment_nodate,dataset_spm_nodate} and to
 the problems formulated by the TEAM (Testing Electromagnetic Analysis Methods) initiative \citep{noauthor_team1_nodate} or the Galileo Ferraris Contest \citep{noauthor_Ferraris_Contest_nodate}, we are not posing a specific challenge.

\section{The Example Case Machines}
The Collaborative Research Centre “Computational Electric Machine Laboratory - CREATOR” (CRC -- TRR361\newline/F90) \citep{noauthor_f90_nodate, noauthor_dfg_nodate}
aims at enhancing electric machine performance and computational efficiency by bringing together advances in many different disciplines. 
As part of this project, 
essential design and measurement data of electric machines are obtained and shall be made available to many other researchers within the community. 
The work that generates the data on these example motors itself assesses the performances of different modeling approaches to analyze dynamic drive cycle operation, including the crucial experimental validation.
The two motors are available and analyzed in the Electric Drives and Power Electronic Systems Institute (EALS) laboratory at TU GRAZ, an induction motor (IM) and a permanent magnet synchronous motor (PMSM). 
While the machines were originally not optimized for traction applications, both are machines designed for research purposes, with all relevant data available and some characteristics that make them notably interesting for research -- and also benchmarking -- purposes. 
The machine design and measurement data packages are made available through two central repositories, \citet{Heidarikani_2024_kh3v7-dhn98.} and \citet{dhakal_2024_sns1d-77m43}, one for each of the two machines. They are discussed and analyzed in Sections \ref{section_IM} and \ref{section_PMSM}, respectively, for each of the two machines.  Prior to this, the next Section outlines the organization of these two main parts of the paper.

\section{Data and Data Presentation Organization}
Subsections \ref{IM_DP} and \ref{PMSM_DP} detail the  repositories of the design data parameters for both the IM and the PMSM, respectively. They provide a structured framework for analysis and comparison. Within these subsections, each part is dedicated to a specific aspect of the motors' designs, as follows: Sub-subsections \ref{IM_geometry}  and \ref{PMSM_Geometry} detail the motor geometries, \ref{IM_Material}  and  \ref{PMSM_Material} the material properties, \ref{IM_electrical}  and  \ref{PMSM_electrical} the electrical parameters and \ref{IM_winding},  and  \ref{PMSM_Widning} the winding schemes of the IM and the PMSM, respectively.

In Subsections \ref{IM_Meas} and \ref{PMSM_Meas}, the repository also provides comprehensive measurement results concerning both the IM's and the PMSM's conventional operating characteristics such as no-load tests for both motors (\ref{PMSM_NL} and \ref{IM_LR&NL}, respectively) and the locked rotor test specifically for the IM \ref{IM_LR&NL}, as well as the derivation of equivalent circuit parameters from measurement data (\ref{IM_Eq} and \ref{PMSM_Eq}, respectively). 

In addition to these steady-state characteristics, the repository provides extensive measurement results of drive cycles, i.e., predefined sequences of torque-speed pairs over time for the analysis of dynamic drive performance. 
Three standard reference drive cycles are selected to evaluate the performance of electric machines in traction electrification: the WLTP (worldwide harmonized light vehicles test procedure), the Braunschweig city drive cycle (urban), and the Artemis 130 km/h drive cycle (inter-city) \citep{dieselnet_emission_nodate}.
The WLTP offers a balanced representation of urban, suburban, and highway driving scenarios. The urban drive cycle emphasizes low-speed, high-torque conditions, crucial for assessing city driving performance characterized by frequent stops and starts. The inter-city drive cycle examines moderate speeds and less frequent stops, reflecting typical conditions encountered during inter-city commutes. 

To derive the required torque and speed for each drive cycle, two reference vehicles are considered: the BMW i3 (medium range) and the Smart EQ (small range), employing their specifications in quasi-static longitudinal vehicle models \citep{Guzzella}. Inputting the drive cycle into the longitudinal vehicle model of each EV determines the required torque and speed on the wheels of the vehicle. The necessary torque and speed of the vehicle motor can then be determined by utilizing the gear transmission. However, since the rating of the vehicle motor may differ from that of the motors in the laboratory, a down-scaling method is employed to ensure that the input data falls within the range of the IM and the PMSM in the laboratory.
Detailed information about the down-scaling method, the longitudinal vehicle model, and vehicle data is presented in \citet{dhakal_down-scaling_2023}.

Schematic overviews of the test benches and the control architectures for both motors are provided in Appendices A and B, respectively. 
 
Sub-subsections \ref{IM_Drive} and \ref{PMSM_Drive} discuss the input drive cycle torque and speed data for the six cases ($3$ cycles, $2$ vehicles) per motor as well as the corresponding measurement results for both machines, including the output and input power and the measured losses of the IM and of the PMSM, respectively.

\section{Induction Motor}\label{section_IM}

The three-phase squirrel cage IM is completely enclosed and water-cooled to maintain optimal performance under varying conditions. With a total of $92$ temperature sensors strategically integrated, this motor enables comprehensive temperature analysis, as it was originally manufactured for analysis of temperature distribution under post-fault operation \citep{eickhoff_space_2021}.

All IM related data and experimental results are made available online through the central repository \citet{Heidarikani_2024_kh3v7-dhn98.}. Table~\ref{general_IM} summarizes the general parameters of the IM.
\begin{table}[h]
\centering
\small
\caption{General design parameters of the IM.}
\label{general_IM}
\begin{tabular}{|l|c|c|}
\hline
\textbf{Parameter}             & \textbf{Value}       &     \textbf{Unit}      \\ \hline

Rated speed                     & $1430$              &              $\si{rpm} $            \\ \hline
Maximum speed                  & $2900$           &                 $\si{rpm} $           \\ \hline

Rated torque & $24.7$               &             $\si{Nm} $            \\ \hline
Maximum torque & $31$          &                $\si{Nm} $            \\ \hline

Rotor moment of inertia  & $0.039$   &            $\si{kg.m^2} $   
\\ \hline
Cooling system                 & Totally enclosed water cooled     &--      \\ \hline
\end{tabular}

\end{table}

\subsection{Design Parameters}\label{IM_DP}
\subsubsection{Motor Geometry}\label{IM_geometry}
\begin{figure}[h!]
\centering
\includegraphics[scale = 0.55]{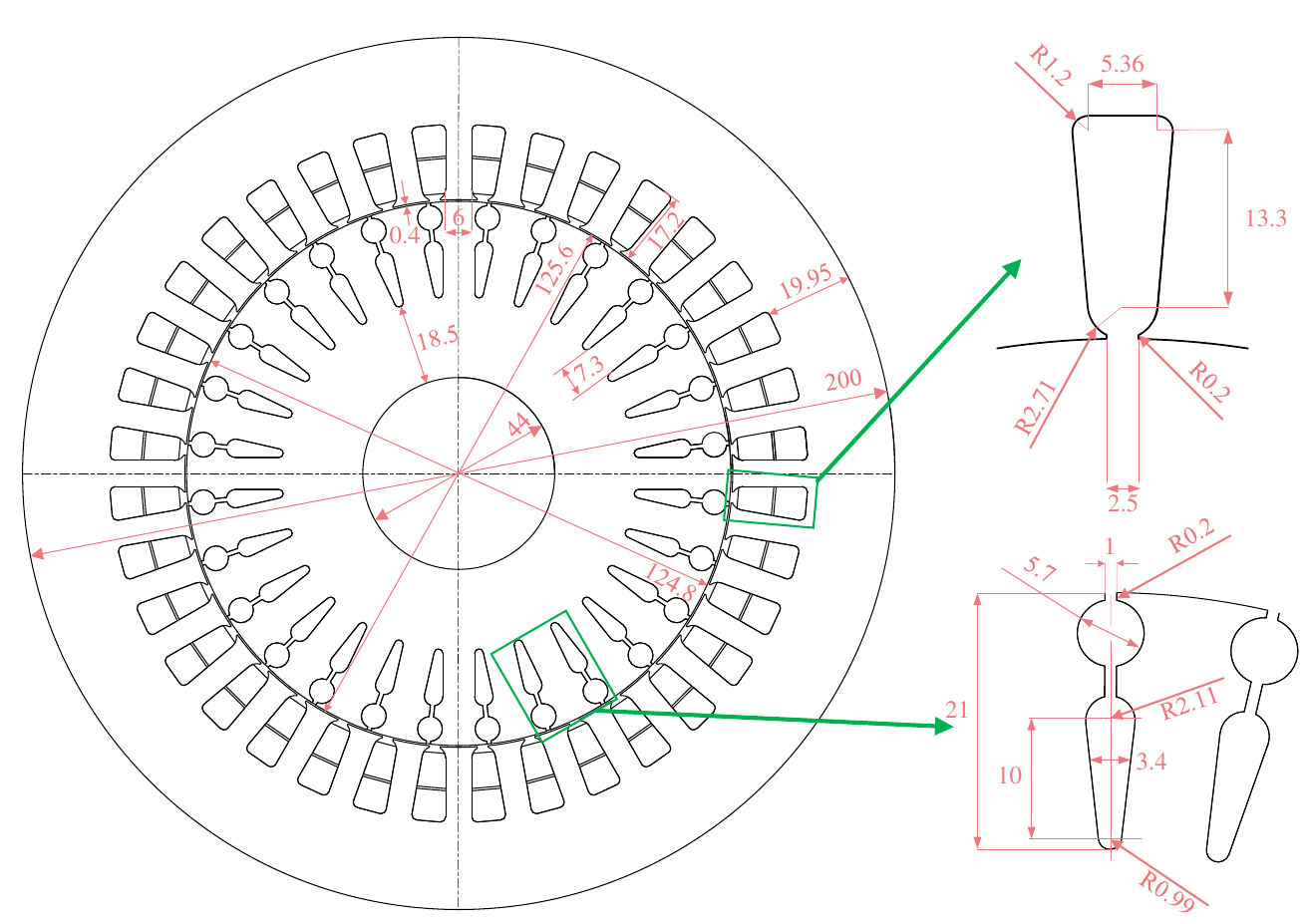}
\caption{2D CAD diagram of the IM with dimensions.}
\label{Geometry}
\end{figure}

Figure~\ref{Geometry} shows a 2D CAD drawing of the motor, with complete dimensions, detailing the motor housing, rotor, stator, and windings. Additional geometric data essential for modeling the IM, such as  details on the end ring, end winding, and housing, are provided in Appendix A.
The online repository
also contains pre-built geometry model files for ease of use. 

\vspace{-2mm}
\subsubsection{Material Properties}\label{IM_Material}
Table~\ref{Mat_IM} shows the materials utilized in the different components of the IM. For each material, the corresponding online data repository link is cited. 
\begin{table}[h!]
\centering
\small
\caption{Material parameters of the IM. }
\label{Mat_IM}
\begin{tabular}{|l|l|}
\hline
\textbf{Part}               & \textbf{Material}           \\ \hline
Stator and rotor iron &   Electrical steel sheet M$800-50$A  \\ \hline
Shaft & Steel \si{1.7225 / 42CrMo4} \\ \hline
Rotor cage and cooling jacket  & Aluminium $6082 / 3.2315$  \\ \hline
Stator winding             &Copper \\ \hline
Slot liners and winding overhang insulation & Trivoltherm NRN  \\ \hline
Stator winding insulation    & Elan-tron MC $4260/$W $4260$ \\ \hline
\end{tabular}
\end{table}
\subsubsection{Electrical Parameters}\label{IM_electrical}
Table~\ref{elec_IM} provides the electrical parameters of the IM. 
\begin{table}[h!]
\centering
\small
\caption{Electrical properties of the IM. }
\label{elec_IM}
\begin{tabular}{|l|c|c|}
\hline
\textbf{Parameter}         & \textbf{Value} & \textbf{Unit} \\ \hline
Maximum power              & $4.6$            & $\si{kW}$        \\ \hline
Nominal power              & $3.7$            & $\si{kW}$        \\ \hline
Nominal voltage            & $400$            & $\si{V.rms,L-L}$ \\ \hline
Maximum dc-link voltage    & $640$            & $\si{V}$         \\ \hline
Nominal current            & $6.9$            & $\si{A.rms}$   \\ \hline
Maximum current            & $15$             & $\si{A.rms}$   \\ \hline
Supply connection          & Y              & --              \\ \hline
Nominal frequency          & $50$             & $\si{Hz}$       \\ \hline
Maximum frequency          & $100$            & $\si{Hz}$       \\ \hline
\end{tabular}
\end{table}

\subsubsection{Winding Scheme}\label{IM_winding}
An overview of the winding scheme is presented here. To this aim, Table~\ref{Wind_IM} details the parameters of the winding of the IM. 
A figure visually illustrating the coil designations for all phases, showing the winding arrangement within the motor is provided in the repository. 

\begin{table}[h!]
\centering
\small
\caption{Winding properties of the IM. }
\label{Wind_IM}
\begin{tabular}{|l|c|c|}
\hline
\textbf{Parameter}                   & \textbf{Value}&     \textbf{Unit}             \\ \hline
Winding type                         & Dual layer distributed winding &  -- \\ \hline

No.\hspace{0.8mm}of phases                  & $3$   &  --                       \\ \hline

Slot per pole and phase              & $3$    &  --                          \\ \hline

No.\hspace{0.8mm}of turns per layer            & $18$   &  --                          \\ \hline
No.\hspace{0.8mm}of stator layer               & $2$      &  --                        \\ \hline

Wire diameter                        &$ 2 \times 0.75 $ &  $\si{mm} $             \\ \hline
Coil pitching                        & $7/9$     &  --                       \\ \hline

\end{tabular}
\end{table}

\subsection{Measurement Results}\label{IM_Meas}
\subsubsection{No-Load and Locked Rotor Tests}\label{IM_LR&NL}
The no-load test evaluates the IM performance without mechanical load. From this, selected parameters such as core loss, magnetization current, and magnetization inductance can be identified.
Conversely, in the locked-rotor test, the motor is mechanically prevented from rotating, and voltage, respectively currents are applied to determine impedance parameters such as rotor resistance and leakage inductances. For more explanations on these tests, see, e.g., \citet{hendershot_design_2010, pyrhonen_design_2009}.

Both tests were carried out for different supply frequencies and amplitudes.

To separate the iron and the sum of friction and windage losses, the measurement results were separated into their frequency-dependent and frequency-independent parts.  The results are also available in the repository.

\subsubsection{Equivalent Circuit Parameters}\label{IM_Eq}
The equivalent circuit parameters of the IM, as derived from the measurements, are shown in  Table~\ref{equi_IM_t}. The stator resistance ($R_{\mathrm{s}}$) was measured through a direct current test, the  rotor resistance ($R_{\mathrm{r}}$) determined from the locked-rotor test. The stator magnetization inductance ($L_{\mathrm{m}}$) was derived from a no-load test, and the combined leakage inductance ($L_{\mathrm{\sigma s}} + L_{\mathrm{\sigma r}}$) was calculated from the locked-rotor test. Due to motor symmetry, the two leakage inductances $L_{\mathrm{\sigma s}}$ and $L_{\mathrm{\sigma r}}$ are assumed to be equal \citep{alberto_leakage}.

\begin{table}[h!]
\centering
\small
\caption{Equivalent circuit parameters of the IM. }
\label{equi_IM_t}
\begin{tabular}{|l|c|c|}
\hline
\textbf{Parameter}                                        & \textbf{Value}     & \textbf{Unit}          \\ \hline
Stator resistance $ R_{\mathrm{s}} $ at 20 $\si{\degree C}$      & $2.329852$           & $\si{\ohm}$            \\ \hline
Rotor resistance $ R_{\mathrm{r}} $ at 20 $\si{\degree C}$ & $1.011977$           & $\si{\ohm}$            \\ \hline
Stator leakage inductance $ L_{\mathrm{\sigma s}} $  & $0.0114450$          & $\si{H}$               \\ \hline
Rotor leakage inductance $ L_{\mathrm{\sigma r}} $    & $0.010648$           & $\si{H}$               \\ \hline
Reference magnetization inductance of stator $ L_{\mathrm{m}} $      & $0.226$              & $\si{H}$               \\ \hline
\end{tabular}
\end{table}

\begin{figure}[h!]
\centering
\includegraphics[scale = 0.60]{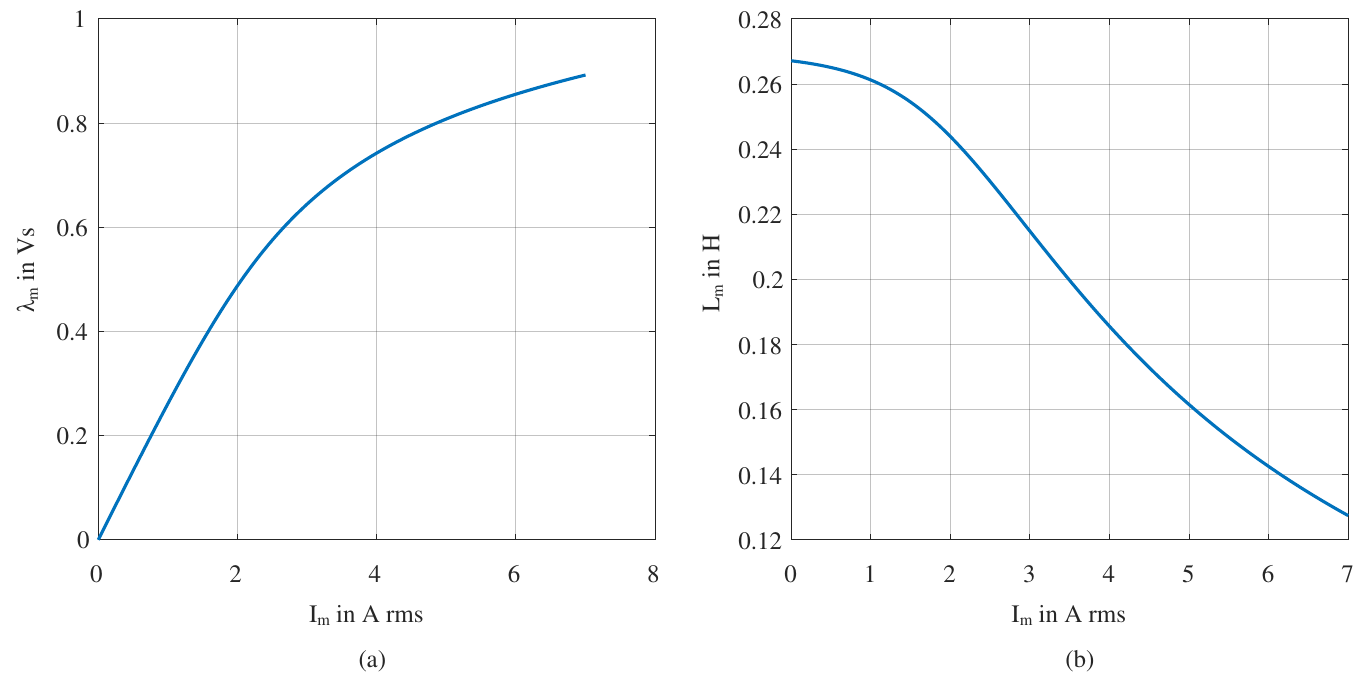}
\caption{Measured magnetization characteristic of the IM: (a) $\lambda_\mathrm{{m}} $ vs $I_\mathrm{{m}}$, (b) $L_\mathrm{{m}} $ vs $I_\mathrm{{m}}$.}
\label{Lm_IM}
\end{figure}

\begin{figure}[h!]
\centering
\includegraphics[scale = 0.60]{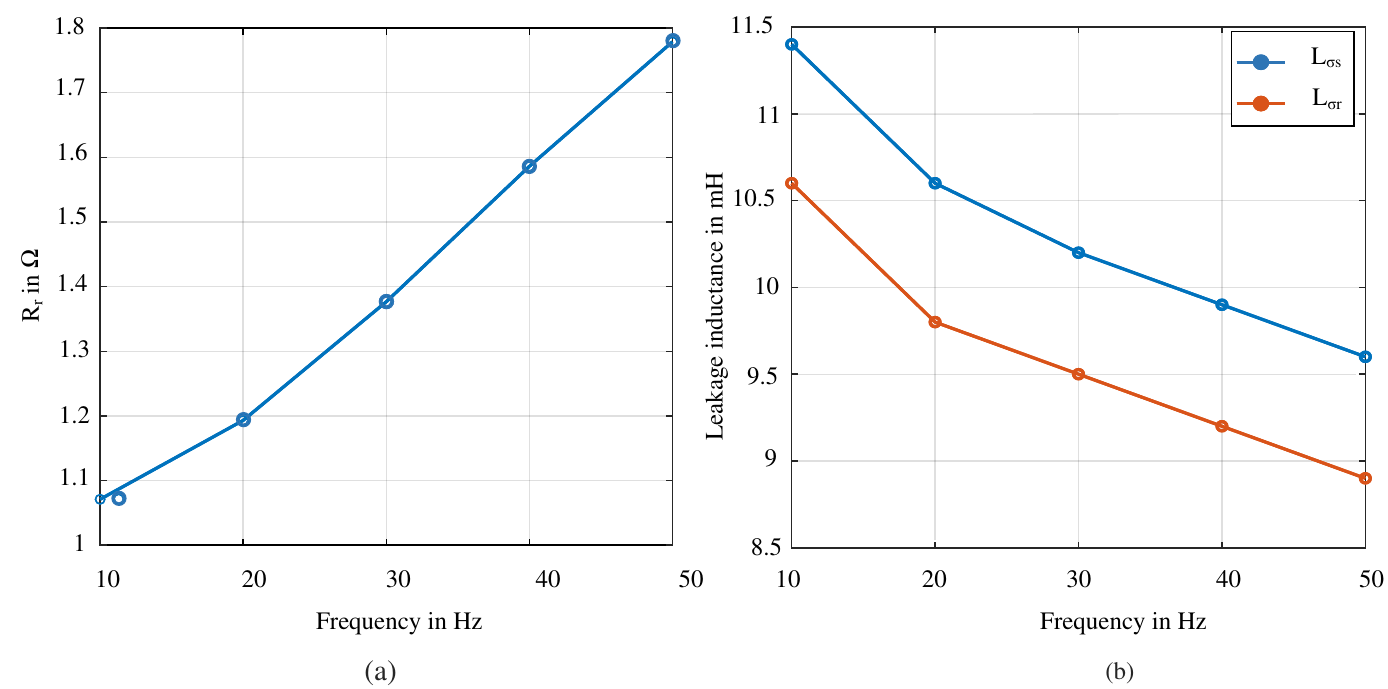}
\caption{Measured rotor resistance and leakage inductances as a function of frequency: (a) $ R_{\mathrm{r}} $, (b) $L_{\mathrm{\sigma s}}$ and $L_{\mathrm{\sigma r}}$. }
\label{Lsig_Rr_Im}
\end{figure}

Figure~\ref{Lm_IM} shows the measured magnetization flux ($\lambda_{\mathrm{{m}}} $) versus current ($I_{\mathrm{{m}}}$) and the derived current-dependent inductance ($L_{\mathrm{ {m}}} $).
Figure~\ref{Lsig_Rr_Im} displays the variation of the rotor resistance with frequency, as well as the relationship between stator and rotor leakage inductances and frequency.
The low-frequency single-phase equivalent circuit of the IM is shown in Appendix A.

\subsubsection{Drive Cycles}\label{IM_Drive}
\begin{figure}[h!]
\centering
\includegraphics[scale = 0.65]{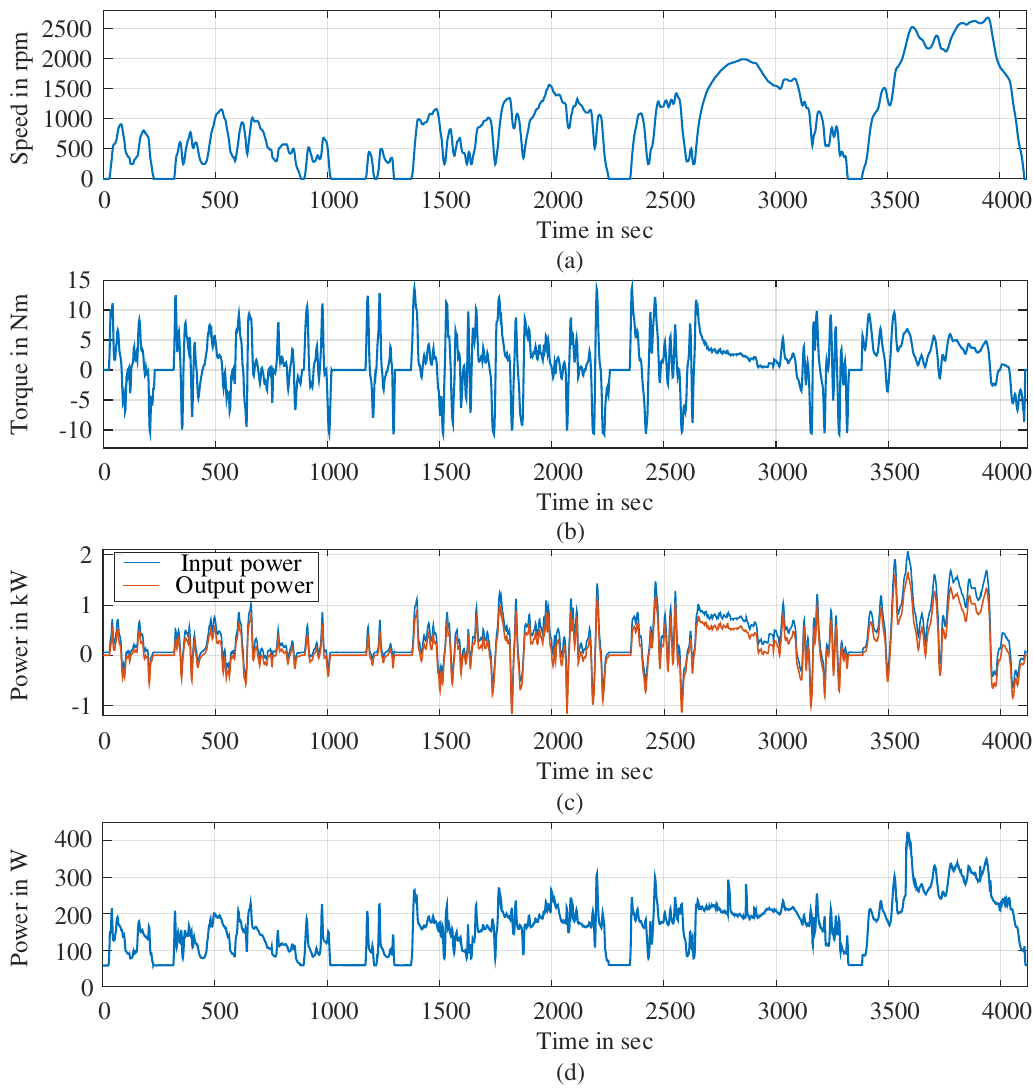}
\caption{Measurement results of WLTP class 3 drive cycle for the IM: (a) speed profile of the down-scaled drive cycle, (b) torque profile of the down-scaled drive cycle, (c) measured input and output powers, (d) calculated total losses.}
\label{in_out_IM}
\end{figure}

An exemplary case of the analysis of the WLTP class $3$ drive cycle for the IM, utilizing the down-scaling procedure \citep{dhakal_down-scaling_2023} based on the specifications of a medium-sized vehicle (BMW i$3$), is presented.
It covers both the input data for the measurement, i.e., the torque and the speed, as well as the measured output data of the IM throughout the entire drive cycle.
The measured input and output powers, along with the total loss calculated from the measurement, as well as the input torque and speed profiles are shown in Figure~\ref{in_out_IM}.
The online repository is not limited to the single drive cycle case depicted in this example; it holds complete measurement data for the six cases of both vehicles and three drive cycles.

\section{PM Synchronous Motor}\label{section_PMSM}
The PMSM machine presented here is a prototype inset PM machine originally designed for use in high-efficiency household cooling appliances with a maximum continuous power of $70\,$W and a targeted optimum performance at $7\,$W. With a maximum continous power and maximum efficiency occurring at different operting points, it reflects demands on many modern motors, notably used in traction applications. 
The prototype machine is naturally cooled and designed to run long hours with considerably high efficiency. All PMSM related data and experimental results are made available online through the central repository \citet{dhakal_2024_sns1d-77m43}. Table~\ref{gen_PMSM} describes the general design parameters of the PMSM. 

\begin{table}[h!]
\small
\centering
\caption{General design parameters of the PMSM.}
\label{gen_PMSM}
\begin{tabular}{|l|c|c|}
\hline
\textbf{Parameter}             & \textbf{Value}   & \textbf{Unit}                 \\ \hline
Rated speed & $2000$ & $\si{rpm}$                      \\ \hline
Maximum speed & $7050$ & $\si{rpm}$                       \\ \hline
Rated torque & $0.10$ & $\si{Nm}$                       \\ \hline
Maximum torque & $0.15$ & $\si{Nm}$                      \\ \hline
Rotor moment of inertia & $0.00011348$ & $\si{kg.m^2} $                    \\ \hline
Cooling system                 & Naturally cooled & -- \\ \hline
\end{tabular}
\end{table}
\subsection{Design Parameters}\label{PMSM_DP}
\subsubsection{Motor Geometry}\label{PMSM_Geometry}
\begin{figure}[ht!]
\centering
\includegraphics[scale = 0.45]{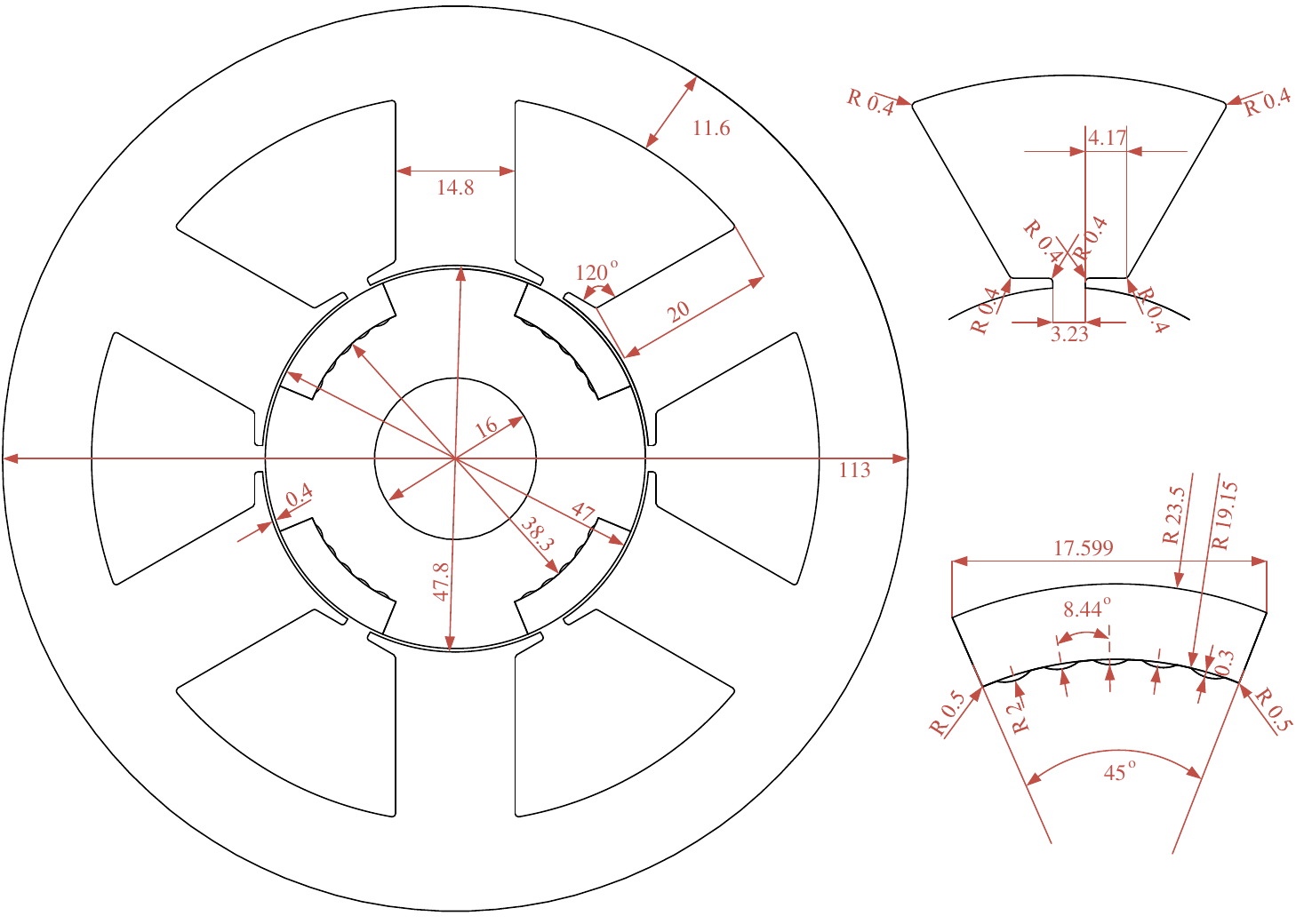}
\caption{2D CAD diagram of the PMSM with  dimensions.}
\label{geometry_PMSM}
\end{figure}

A 2D geometry diagram of the PMSM is illustrated in Figure~\ref{geometry_PMSM}. The stator, rotor, and magnet dimensions are highlighted. Table~\ref{Geo_PMSM} details the motor geometry. For ease of use, pre-built geometry model files are also available in the online repository.

\subsubsection{Motor Material} \label{PMSM_Material}
Table~\ref{materialsPMSM} details the motor parts and the corresponding material used. For each material, the corresponding online repository link is cited. 
\begin{table}[h!]
\small
\centering
\caption{Material parameters of the PMSM. }
\label{materialsPMSM}
\begin{tabular}{|l|c|}
\hline
\textbf{Part}    & \textbf{Material} \\ \hline
Stator and rotor iron              & Electrical steel M$250-35$A                          \\ \hline
Stator winding              & Copper                           \\ \hline
Magnet              & Sintered ferrite                           \\ 
                    & radially magnetized       \\ \hline
Shaft              & --                          \\ \hline
\end{tabular}
\end{table}
\subsubsection{Electrical Parameters}\label{PMSM_electrical}
Table~\ref{elecPMSM} provides the electrical parameters of the PMSM.
\begin{table}[h!]
\small
\small
\centering
\caption{Electrical parameters of the PMSM. }
\label{elecPMSM}
\begin{tabular}{|l|c|c|}
\hline
\textbf{Parameter}    & \textbf{Value}  & \textbf{Unit} \\ \hline
Maximum power             & $70$ & $\si{W}$                           \\ \hline
Optimum power             & $7$ & $\si{W}$                          \\ \hline
Nominal voltage              & $135$ & $\si{V.rms}$                          \\ \hline
Maximum dc-link voltage              & $326$ & $\si{V}$                          \\ \hline
Nominal current              & $0.21$ & $\si{A.rms}$                          \\ \hline
Maximum current              & $0.3$ & $\si{A.rms}$                         \\ \hline
Supply connection              & Y  & --                        \\ \hline
Nominal frequency             & $66.67$ & $\si{Hz}$                         \\ \hline
Maximum frequency             & $235$ & $\si{Hz}$                          \\ \hline
\end{tabular}
\end{table}
\subsubsection{Winding Scheme}\label{PMSM_Widning}
Here, the winding scheme of the PMSM is presented. Table~\ref{wind_PMSM} details the winding properties.
A figure showing its model and the slot view is provided in the repository.

\begin{table}[h!]
\small
\centering
\caption{Winding properties of the PMSM. }
\label{wind_PMSM}
\begin{tabular}{|l|c|c|}
\hline
\textbf{Parameter}                   & \textbf{Value} & \textbf{Unit}                  \\ \hline
Winding type       & Tooth wound/concentrated winding & --           \\ \hline
No.\hspace{0.8mm}of phases                  & $3$ & --                         \\ \hline
Slot per pole per phase              & $0.5$ & --                             \\ \hline
No.\hspace{0.8mm}of turns per slot  & $328$ & --                      \\ \hline
No.\hspace{0.8mm}of winding layer               & $1$ & --                              \\ \hline
Copper wire diameter              & $0.64$ & $\si{mm}$                          \\ \hline
\end{tabular}
\end{table}

\subsection{Measurement results}\label{PMSM_Meas}
\subsubsection{No-Load Tests}\label{PMSM_NL}
\begin{figure}[h!]
\centering
\includegraphics[scale = 0.60]{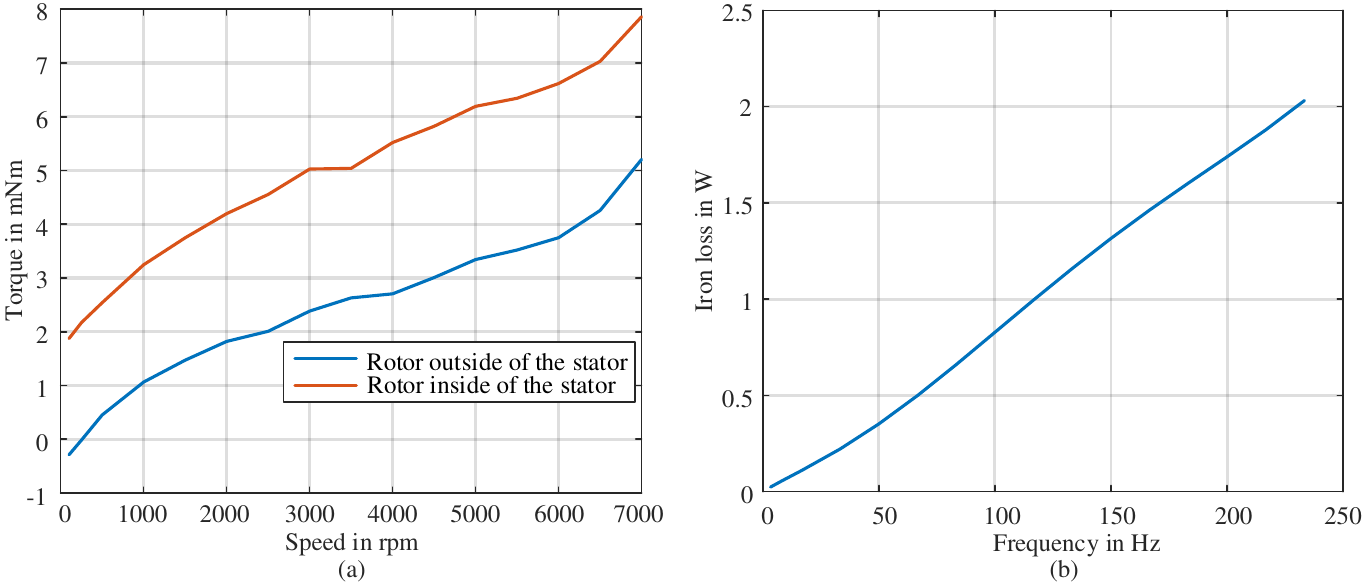}
\caption{No-load measurement results of the PMSM: (a) torque for rotor inside and outside of the stator, (b) calculated iron losses.}
\label{NL_PMSM}
\end{figure}

No-load tests of the PMSM are carried out to assess the load-independent losses within the motor itself, such as iron and frictional losses, and to assess  the machine parameters, such as back-EMF and cogging torque. For more explanations on such kinds of tests, see, e.g., \citet{hendershot_design_2010, pyrhonen_design_2009}. 
The PMSM is tested under no-load, both with the rotor mounted within the stator and without rotor (to identify the iron loss), at different rotor speeds. 
Figure~\ref{NL_PMSM} shows the no-load torque measurement results as function of speed and the calculated no-load iron losses as function of supply frequency.

For the measurement of the cogging torque, the rotor is rotated very slowly, at $0.25\,$rpm. The torque measurements are taken by a torque transducer. To measure the back-EMF, the rotor was rotated at a constant speed of $2000\,$rpm.
The results are also provided in the repository.
A fundamental peak component of the induced back-EMF of $47.5\,$V per phase is identified by discrete Fourier transform.

\subsubsection{Equivalent Circuit Parameters}\label{PMSM_Eq}
Table~\ref{equiPMSM} shows the low-frequency equivalent circuit parameters of the PMSM. The stator resistance is measured directly using an LCR meter \citep{lcr_meter}. The $d$-axis and the $q$-axis inductances are obtained through finite element analysis (FEA) using the JMAG$^{\circledR}$ software's \citep{noauthor_simulation_nodate} inductance calculator tool \citep{jmag_inductance}. An exemplary 2D JMAG model file used for this analysis is also made available in the repository.´
\begin{table}[h!]
\small
\centering
\caption{Equivalent circuit parameters of the PMSM.}
\label{equiPMSM}
\begin{tabular}{|l|c|c|}
\hline
\textbf{Parameter}                    & \textbf{Value} & \textbf{Unit}        \\ \hline
Fundamental back emf at $2000$ $\si{rpm}$, $E_{\si{0}}$       & $47.37$ & $\si{V.peak}$           \\ \hline
Cogging torque at $0.25$ $\si{rpm}$        & $0.0357$ & $\si{Nm.peak}$            \\ \hline
Magnet flux linkage, $\lambda_{\si{pm}}$        & $0.1144$ & $\si{Wb}$            \\ \hline
Stator inductance d-axis, $L_{\si{d}}$        & $0.2055$ & $\si{H}$            \\ \hline
Stator inductance q-axis, $L_{\si{q}}$         & $0.3320$ & $\si{H}$            \\ \hline
Stator phase resistance, $R_{\si{s}}$       & $8.9462$ & $\si{\ohm}$           \\ \hline
\end{tabular}
\end{table}

To calculate the current angle for maximum torque, the current angle was varied from $90$\degree$_\text{el.}$ to $120$\degree$_\text{el.}$, with current amplitudes of $0.15\,$A and $0.3\,$A. The motor was set to operate under motoring operating points at a constant speed of $2000\,$rpm. The torque was measured as the current angle was varied and the angle that gives the maximum torque value is identified.
The maximum current angles with current amplitude of $0.15\,$A and $0.3\,$A are found to be $100$\degree$_\text{el.}$ and $110$\degree$_\text{el.}$, respectively.
 The low frequency single phase equivalent circuit models of the PMSM are provided in Appendix B. 

\subsubsection{Drive Cycles}\label{PMSM_Drive}
\begin{figure}[h!]
\centering
\includegraphics[scale = 0.65]{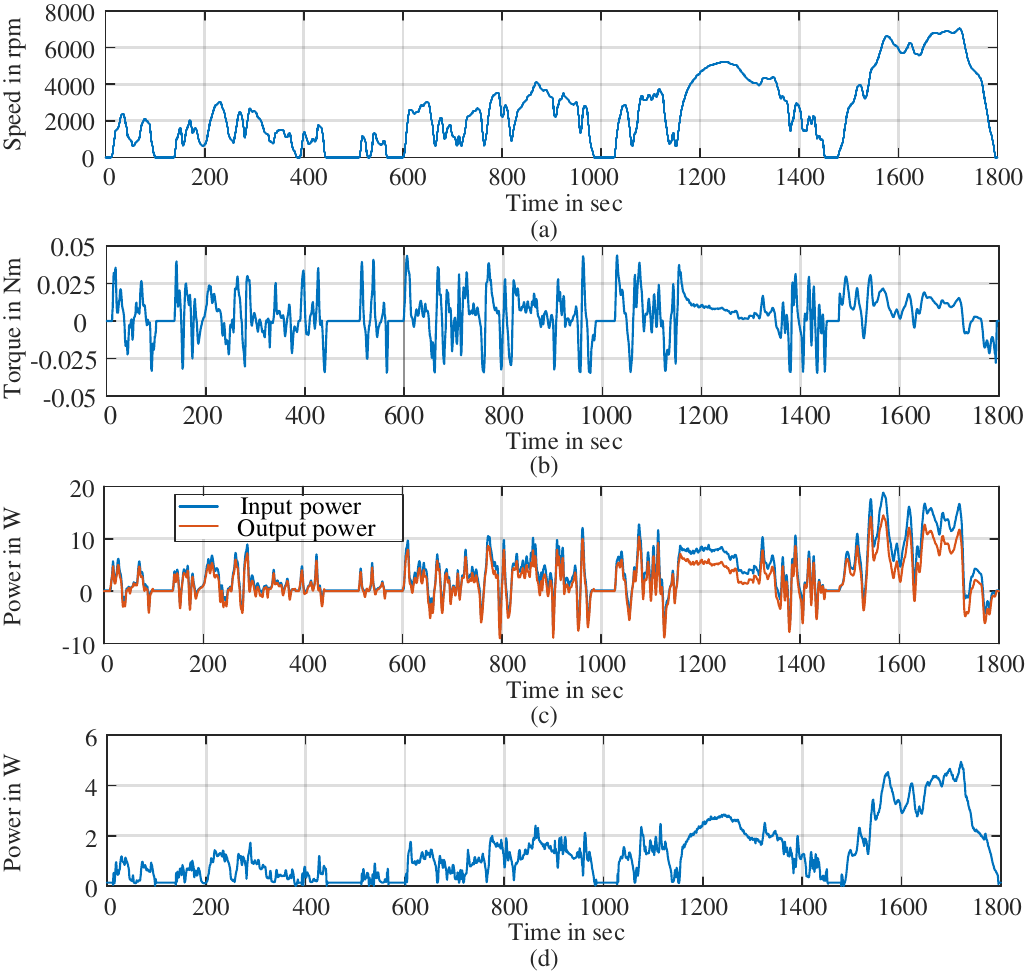}
\caption{Measurement results of WLTP class 3 drive cycle for the PMSM: (a) speed profile of the down-scaled drive cycle, (b) torque profile of the down-scaled drive cycle, (c) measured input and output powers, (d) calculated total losses.}
\label{measPMSM}
\end{figure}

Exemplarily, the measurement results of the WLTP class 3 drive cycle with reference to the medium-sized passenger car (BMW i$3$), down-scaled to the PMSM as per \citet{dhakal_down-scaling_2023}, are shown here, see Figure~\ref{measPMSM}. They include the drive cycle input speed and torque, the input power, the output power, and the total losses throughout the whole drive cycle.
The online repository contains the complete set of measurement data for both vehicles and all three drive cycles. 

\section{Conclusion}
This paper presents comprehensive electric machine design parameters and measurement results of two different electric machines and for a large set of six drive cycles per machine. It makes all relevant design data available for benchmarking of modeling and simulation approaches. 
This paper not only describes the data itself but also guides the prospective user through their organization within the repository.

\section*{Acknowledgment}
This work is partially supported by the joint Collaborative Research Centre CREATOR (DFG: Project-ID 492661287/TRR 361; FWF: 10.55776/F90) at TU Darmstadt, TU Graz and JKU Linz.
The authors would like to thank Dr. Hermann Schranzhofer from TU Graz for his help with the realization of the repositories as well as helpful suggestions on the overall organization of the open science contribution of these data. They would also like to thank Mario Mally and Michael Wiesheu from TU Darmstadt for through proofreading of an earlier version of this paper.

\section*{Appendix A } \label{appendix_IM}
\subsection*{Geometry of the IM}
The detailed geometric parameters of the IM are listed in Table~\ref{Geo_IM}.
\begin{table}[h]
\small
\centering
\caption{Main geometry parameters of the IM. }
\label{Geo_IM}
\begin{tabular}{|l|c|c|}
\hline
\textbf{Parameter}                                 & \textbf{Value} & \textbf{Unit} \\ \hline

Thickness of housing                               & $0.015$          & $\si{m}$       \\ \hline

Length of housing                                  & $0.23$           & $\si{m}$       \\ \hline

Stator iron inner diameter                         & $0.1256$         & $\si{m}$       \\ \hline

Stator iron outer diameter                         & $0.2$            & $\si{m}$       \\ \hline

Length of stator iron                              & $0.1$            & $\si{m}$       \\ \hline

Length of stator teeth                             & $0.0172$         & $\si{m}$       \\ \hline

Width of stator teeth \si{(equivalent)}           & $0.0062$         & $\si{m}$       \\ \hline

Stator slot opening                                & $0.0025$         & $\si{m}$       \\ \hline

Average wire radius                                & $0.0031$         & $\si{m}$       \\ \hline

Stator end winding length                          & $0.03$           & $\si{m}$       \\ \hline

Stator end winding inner diameter                  & $0.135$          & $\si{m}$       \\ \hline

Stator end winding outer diameter                  & $0.18$           & $\si{m}$       \\ \hline

Stator slot fill factor                                 & $30$             & \%             \\ \hline

No.\hspace{0.8mm}of stator slots                   & $36$             &    --           \\ \hline

Rotor core inner diameter                          & $0.044$          & $\si{m}$       \\ \hline

Rotor core outer diameter                          & $0.1248$         & $\si{m}$       \\ \hline

Rotor end ring length \si{(axial)}                      & $0.014$          & $\si{m}$       \\ \hline

Rotor end ring inner diameter                      & $0.081$          & $\si{m}$       \\ \hline

Rotor end ring outer diameter                      & $0.1242$         & $\si{m}$       \\ \hline

Rotor slot average width                           & $0.0034$         & $\si{m}$       \\ \hline

Rotor slot average height                          & $0.021$          & $\si{m}$       \\ \hline

No.\hspace{0.8mm}of rotor cage bars                & $28$             &    --           \\ \hline

Motor air gap                                      & $0.0004$         & $\si{m}$       \\ \hline

No.\hspace{0.8mm}of poles                          & $4$              &    --           \\ \hline
\end{tabular}
\end{table}

\subsection*{Equivalent Circuit Model of the IM}
\begin{figure}[h!]
\centering
\includegraphics[scale = 0.5]{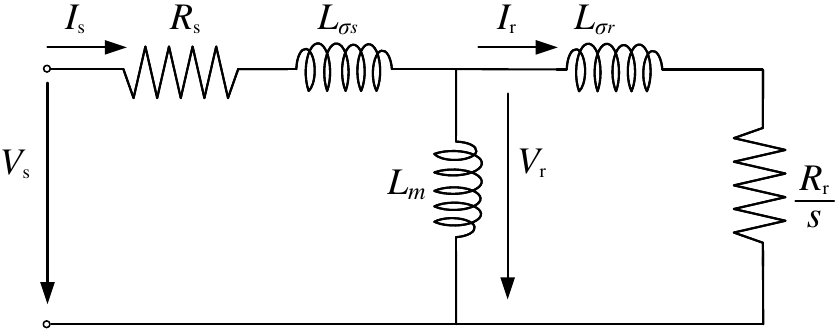}
\caption{Single phase ECM of an IM.}
\label{equi_IM}
\end{figure}

As is common for simplicity, the IEEE single-phase equivalent circuit model (ECM) of the IM is presented without considering core losses, as shown in Figure~\ref{equi_IM} \citep{hengameh_equivalet_IM}. The diagram illustrates the per-phase equivalent circuit of the IM, where the rotor-side impedances are converted to the stator side. In the circuit, $R_s$ and $R_r$ represent the stator and rotor resistances (referred to the stator side), while $L_{\mathrm{\sigma s}}$ and $L_{\mathrm{\sigma r}'}$ denote the stator and rotor leakage inductances. The magnetization inductance is represented by $L_m$. Additionally, the stator voltage $V_s$, rotor voltage $V_r$, stator current $I_s$, and rotor current $I_r$ are depicted. The values of the equivalent circuit parameters for the IM can be found in Table~\ref{equi_IM_t}. 

\subsection*{Laboratory Test Setup and Control Architecture of the IM}
\begin{figure}[h!]
\centering
\includegraphics[scale = 0.65]{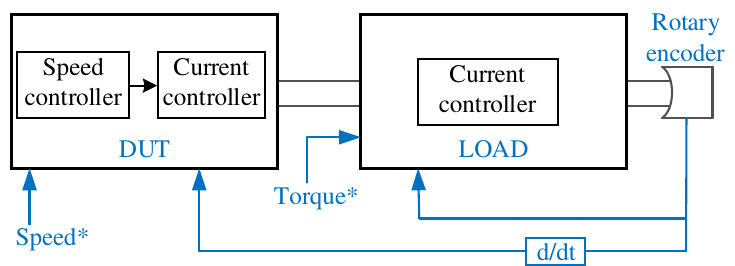}
\caption{Control scheme for the drive cycle of the IM on the laboratory test bench, where * denotes the commanded values.}
\label{control_IM}
\end{figure}

Rotor field-oriented control (RFOC) is utilized to control the speed of the IM, here the device under test (DUT). Additionally, the torque control is applied to the PMSM, which functions as the load machine.
The control architecture illustrated in Figure~\ref{control_IM} outlines the comprehensive control scheme utilized in the laboratory's IM test bench. Both inverters use space vector pulse width modulation (SVPWM), at a constant switching frequency of $5\,$kHz.

\section*{Appendix B} \label{appendix_PMSM}
\subsection*{Geometry of the PMSM}
Table~\ref{Geo_PMSM} lists the detailed geometry parameters of the PMSM. 
\begin{table}[h!]
\small
\centering
\caption{Main geometry parameters of the PMSM. }
\label{Geo_PMSM}
\begin{tabular}{|l|c|c|}
\hline
\textbf{Parameter}    & \textbf{Value} & \textbf{Unit} \\ \hline
Stator outer diameter              & $113$ &$\si{mm}$                          \\ \hline
Stator inner diameter              & $47.8$ &$\si{mm}$                          \\ \hline
Width of back iron              & $11.6$ &$\si{mm}$                           \\ \hline
Slot height \si{(equivalent)}             & $19.6$ &$\si{mm}$                           \\ \hline
Slot width \si{(equivalent)}               & $21.6$ &$\si{mm}$                           \\ \hline
Width of slot opening              & $3.23$ &$\si{mm}$                           \\ \hline
Height of slot opening              & $0.98$ &$\si{mm}$                           \\ \hline
Stator slot fill factor              & $0.5$ & --                          \\ \hline
Tooth height \si{(equivalent)}               & $20$ &$\si{mm}$                           \\ \hline
Tooth width \si{(equivalent)}               & $14.8$ &$\si{mm}$                           \\ \hline
Rotor outer diameter              & $47$ &$\si{mm}$                           \\ \hline
Air-gap length              & $0.4$ &$\si{mm}$                           \\ \hline
Magnet length \si{(radial\ direction)}               & $4.35$ &$\si{mm}$                           \\ \hline
Magnet height              & $17.6$ &$\si{mm}$                           \\ \hline
Magnet span              & $45$ &$\si{\degree.mechanical}$                        \\ \hline
Shaft diameter              & $16$ &$\si{mm}$                           \\ \hline
Motor length              & $30.1$ &$\si{mm}$                           \\ \hline
No.\hspace{0.8mm}of poles              & $4$ & --                          \\ \hline
No.\hspace{0.8mm}of stator slots                     & $6$  & --                       \\ \hline
Stator end winding length              & $11.4$ &$\si{mm}$                           \\ \hline
Thickness of stator and rotor lamination              & $0.35$ &$\si{mm}$                           \\ \hline
\end{tabular}
\end{table}

\subsection*{Equivalent Circuit Model of the PMSM}

\begin{figure}[t!]
\centering
\includegraphics[scale = 0.5]{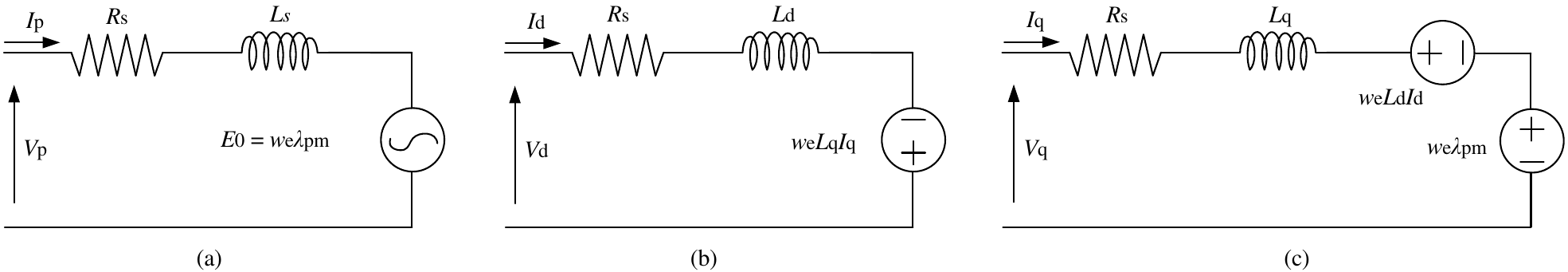}
\caption{ECMs of a PMSM: (a) single phase ECM, (b) d-axis ECM, (c) q-axis ECM.}
\label{equi_PMSM}
\end{figure}

In the majority of analyzed PMSM cases, single-phase ECMs are presented, disregarding the core loss component \citep{ba_development_2022}. Simplified ECMs of a PMSM are demonstrated in Figure~\ref{equi_PMSM}. Figure~\ref{equi_PMSM}(a) presents the single phase ECM of a PMSM. As presented in the diagram, $R_{\si{s}}$ is the per-phase resistance of the stator winding and $L_{\si{s}}$ is the synchronous inductance, which is an equivalent inductance of self and mutual per-phase inductances. The flux linkage of permanent magnets is denoted by $\lambda_{\si{pm}}$. The back electromotive force, $E_{\si{0}}$ is proportional to the electrical rotational frequency $\omega_{\si{e}}$. The single phase current and voltage are denoted by $I_{\si{p}}$ and $V_{\si{p}}$, respectively. Figure~\ref{equi_PMSM}(b, c), represent the corresponding d and q-axis ECMs, respectively. As denoted in the diagram, $V_{\si{d}}$ and $V_{\si{q}}$ are the d nad q-axis terminal voltages, and $I_{\si{d}}$ and $I_{\si{q}}$ are the d and q-axis armature currents. The d and q-axis inductances are denoted by $L_{\si{d}}$ and $L_{\si{q}}$, respectively. The corresponding equivalent circuit parameter values of the PMSM can be refereed from Table~\ref{equiPMSM}.
\subsection*{Laboratory Test Setup and Control Architecture of the PMSM}

\begin{figure}[h!]
\centering
\includegraphics[scale = 0.65]{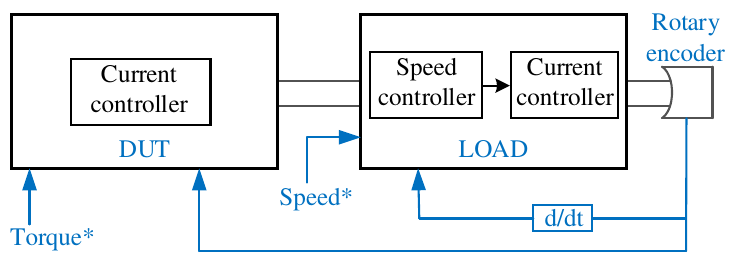}
\caption{Drive cycle control scheme of the PMSM in the laboratory test bench, * denotes that the value is commanded.}
\label{control_PMSM}
\end{figure}

To control the drive cycle's input speed and torque, a cascaded control technique is employed. Figure~\ref{control_PMSM} illustrates the overall control scheme of the PMSM test bench in the laboratory. The DUT, which is the PMSM presented here, is torque-controlled, and the load machine, which is also a PMSM, is speed-controlled. The choice to torque-control the DUT was based on the goal of obtaining fewer harmonics in the input currents at the machine terminal. Both machines side inverters operate at a switching frequency of $20\,$kHz. A maximum torque per ampere (MTPA) control scheme is used for the current controllers.

\end{document}